\documentstyle[budapest1]{article}  
\frompage{261} \topage{268}                                              
\def\rightmark{Hadronic Spectral Function ...} 
\def\leftmark{D. Blaschke et al.}
 
\title{Hadronic Spectral Function and Charm Meson Production} 
\authors{
{D. Blaschke$^{1,2}$, G. Burau$^1$, T. Barnes$^{3,4}$, Yu. Kalinovsky$^{5}$
\\and E. Swanson$^{6}$
}\\[2.812mm]
{\normalsize
\hspace*{-8pt}$^1$ Fachbereich Physik, Universit\"at Rostock\\ 
18051 Rostock, Germany\\[0.2ex] 
\hspace*{-8pt}$^2$ Bogoliubov Laboratory for Theoretical Physics,
JINR Dubna\\ 
141980 Dubna, Russia\\[0.2ex]
\hspace*{-8pt}$^3$ Physics Division, Oak Ridge National Laboratory,\\ 
Oak Ridge, TN 37831, USA\\[0.2ex]
\hspace*{-8pt}$^4$ Department of Physics, University of Tennessee,\\ 
Knoxville, TN 37996, USA\\[0.2ex]
\hspace*{-8pt}$^5$ Laboratory for Information Technologies, JINR Dubna\\ 
141980 Dubna, Russia\\[0.2ex]
\hspace*{-8pt}$^6$ Department of Physics and Astronomy, University of 
Pittsburgh,\\ Pittsburgh, PA 15260, USA\\[0.2ex]
}}
 
\abstract{At the chiral restoration/deconfinement transition, most hadrons
undergo a Mott transition from being bound states in the confined phase to
resonances in the deconfined phase. We investigate the consequences of
this qualitative change in the hadron spectrum on 
final state interactions of charmonium in hot and dense matter, and show that 
the Mott effect for D-mesons leads to 
a critical enhancement of the J/$\psi$ 
dissociation rate. Anomalous J/$\psi$ suppression in the NA50 experiment is
discussed as well as the role of the Mott effect for the heavy flavor kinetics 
in future experiments at the LHC. The status of our calculations of
hadron-hadron cross sections using the quark interchange and chiral Lagrangian 
approaches is reviewed, and an Ansatz for a unification of these schemes
is given.}

 
\begin{document}
 
\maketitle
\setcounter{page}{1}

\section{Introduction}\label{intro}
 
Charmonium states, in particular the J/$\psi$ meson, play a key role in the 
experimental search for the quark-gluon plasma (QGP) in ultrarelativistic 
heavy-ion collisions. The anomalous suppression of J/$\psi$ production 
found in 158 AGeV Pb-Pb collisions at the CERN SPS by the NA50 collaboration 
\cite{na50} seems
to be the signal of QGP formation, as 
originally suggested by Matsui and Satz
\cite{ms86}. 
Since the first observation of J/$\psi$ suppression in nucleus-nucleus 
collisions by the NA38 collaboration \cite{na38} a debate has been started
whether this observation proves QGP formation or whether it can be understood
in terms of more conventional hadronic absorption mechanisms on projectile/
target nucleons \cite{gh88} and comoving hadrons formed in the collision 
\cite{v88,gg88}. So, there is a need for
a refinement and extension of
the experimental information as well as for a unifying theoretical approach 
which can consistently account for the complexity of the processes in
high-energy hadronic collisions in a description based on quark and
gluon substructure, including aspects of the QGP phase transition.
 
In the present contribution we will consider the 
characteristic energy dependence of the J/$\psi$ dissociation cross
section in collision with light hadrons in the constituent quark model,
as well as considering their dissociation into 
$q\bar q$ resonances at the chiral/deconfinement phase 
transition (Mott effect). 

\section{Quark substructure effects in meson-meson scattering} 
\label{scattering}
\begin{figure}[ht]
                 \insertplot{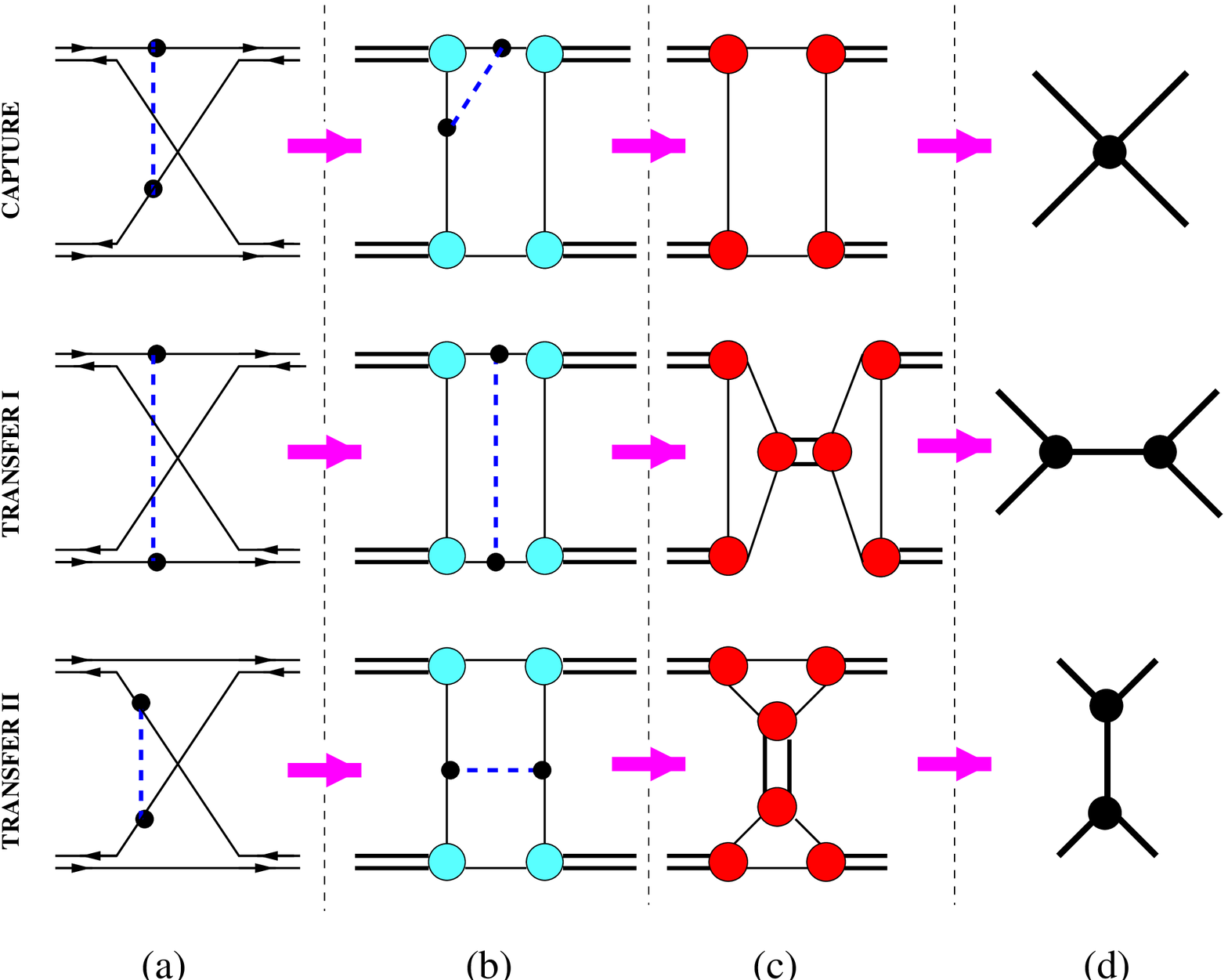}
\caption[]{Diagrammatic scheme of the relation between the quark interchange 
model (a) and the chiral Lagrangian approach (d) of meson-meson scattering.
Both approaches can be viewed as limits of a more general chiral quark
model approach (c). The transition (c) $\to$ (d) is a local limit in which
the meson-quark-antiquark vertices are replaced by coupling constants and the 
momentum dependence of the quark propagators is dropped. The transition (c)
$\to$ (a) is for the two lower lines (transfer diagrams) a replacement of the 
meson propagator by a single interaction (1st Born approximation) and for the 
first line (capture diagram) a first iteration of the meson-quark-antiquark
vertex function using the Bethe-Salpeter equation.}
\label{david}
\end{figure}

\begin{figure}[ht]
                 \insertplot{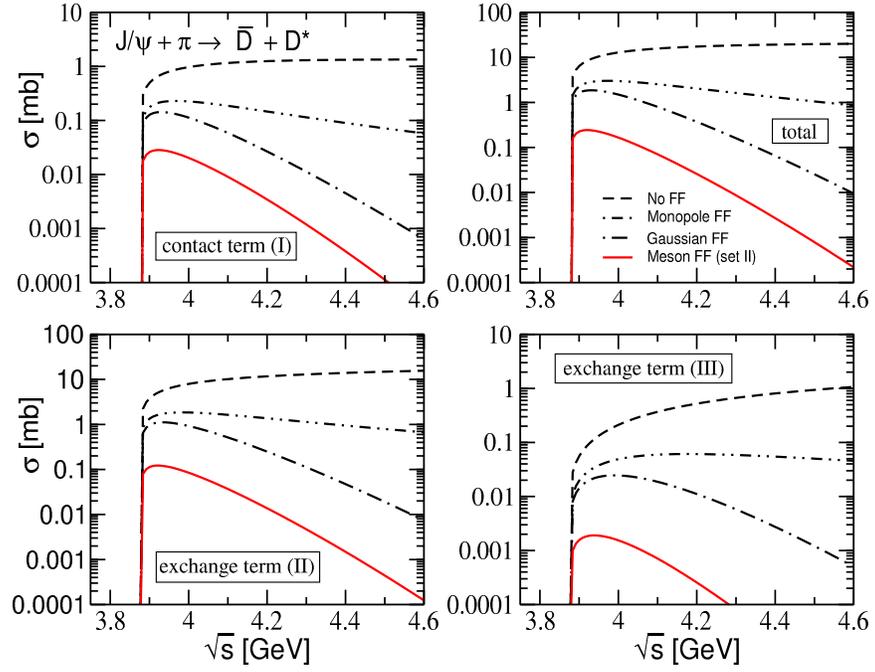}
\caption[]{Cross section for charmonium dissociation in the chiral Lagrangian 
approch with mesonic form factors \cite{bbk01} compared to the global 
form factors of Refs. \cite{lk00,hg00}.}
\label{jpsipi_Ib}
\end{figure}

\begin{figure}[ht]
                 \insertplot{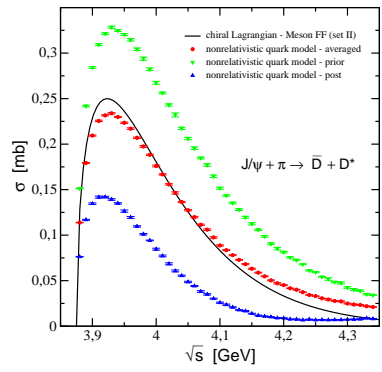}
\caption[]{Comparison of cross sections for chiral Lagrangian model 
with mesonic form factors \cite{bbk01} and quark interchange model 
\cite{wsb00}.}
\label{jpsipi_II_neu}
\end{figure}

In Fig. \ref{david} we illustrate diagrammatically the relation between
(a) the quark interchange model (QIM) and (d) the chiral Lagrangian model
(CLM) of meson-meson scattering. Both approaches can be considered as
limiting cases of a more general chiral quark model (CQM) approach (c).
After the first calculation of J/$\psi$ dissociation due to pion and rho-meson
impact within the CLM \cite{mm98}, several improvements
have been suggested, in particular the use of a global form factor in order
to account for the finite size of the interaction vertex  
\cite{lk00,hg00,Oh:2000qr}. 
From the point of view of the unifying CQM \cite{b+00}, however, the contact 
terms and the meson exchange terms of the CLM should not
have identical form factors since they are composed of a different number
of corresponding meson-quark-antiquark vertices so that the relative 
importance of these subprocesses will depend on the actual parametrization
\cite{bbk01,Blaschke:2001}. It is unsatisfactory that these approaches depend
strongly on the rather arbitrary definition of the form factor, see Fig. 
\ref{jpsipi_Ib} for a comparison. 
The correspondence of CLM and QIM \cite{mbq95,wsb00} results for the J/$\psi$
dissociation cross section, Fig. \ref{jpsipi_II_neu} and Refs. 
\cite{bbk01,Blaschke:2001,Oh:2002}, which can be considered as a heuristic 
constraint on the choice of the form factors, is yet accidental. 
Although the dependence of $\sigma(s;M_{D_1},M_{D_2})$ on the $D$-meson masses 
in both approaches is qualitatively similar and will be important for our 
definition of the in-medium dissociation 
cross section, the dependence on the size 
parameters $\Lambda_h^{-1} \propto \sqrt{\langle r^2\rangle_h}$ of the mesons 
predicted by the two approaches differ.
Whereas the QIM cross section approximately reproduces \cite{mbq95} the 
phenomenological Povh-H\"ufner law \cite{ph} 
$\sigma \propto \langle r^2\rangle_{h_1}\langle r^2\rangle_{h_2}$
for the processes considered here,
the CLM cross section does not. This discrepancy may indicate
the necessity of including hadronic form factors in the CLM;
this possibility is currently under investigation
\cite{bbk02}.       


\section{Spectral properties of mesonic states at finite T}

\begin{figure}[ht]
                 \insertplot{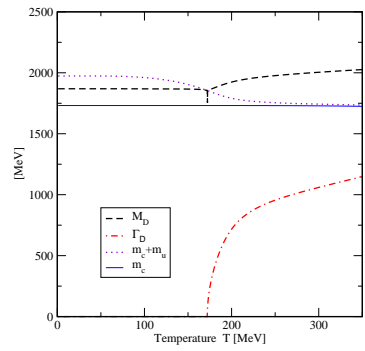}
\caption[]{D-meson mass and width at finite temperature from the NJL model
with infrared cutoff of Ref. \cite{bbvy}.}
\label{DmesonNJL}
\end{figure}

Mesons are not elementary objects. 
In some models light mesons such as the
$\rho$ and the light ``$\sigma$-meson'' effect can be viewed either as 
quark-antiquark bound states or as resonances 
of the $\pi-\pi$ interaction in the corresponding channel. 
The total spectral width $\Gamma_{\sigma}(T)$ of the $\sigma$-meson, e.g., 
shows a minimum correlated with the chiral restoration phase transition in 
the phase diagram of strongly interacting matter \cite{kv+98} since the 
hadronic decay width $\Gamma_{\sigma \to 2 \pi}$ is already negligible but 
the decay width $\Gamma_{\sigma \to q \bar q}$ is still small.  
The transition from a bound state with vanishing decay width (infinite 
lifetime) to a resonance in the continuum of unbound states is called
the Mott transition \cite{rr} and can be described by the behaviour of the 
spectral function
\begin{equation}
\label{a}
A_h(s;T)=\frac{1}{N}\frac{\Gamma_h(T)~M_h(T)}
{[s-M_h^2(T)]^2+\Gamma^2_h(T)~M^2_h(T)}~,
\end{equation}
where $\Gamma_h(T)$ and $M_h(T)$ are the temperature dependent width and mass
parameter of the hadron $h$. 
An introduction into critical phenomena related to the Mott transition for 
mesons at the chiral/ deconfinement transition within the NJL model for 
quark matter can be found in \cite{hkr96}.
It has been found within this model that the Mott transition temperature
for $D$ mesons almost coincides with that for $\pi$ and $K$ mesons
\cite{gk92}. In order to obtain the parameters of the spectral function 
(\ref{a}) for the light and charm mesons, we use a modified NJL model with 
infrared confinement (no unphysical quark production thresholds) \cite{bbvy},
see Fig. \ref{DmesonNJL}, and compare to a fit with linear $T$-dependence
of mass and width \cite{bbk2} above the Mott temperature. 
In the following, we investigate
the consequences of the meson Mott effect for charmonium production.

 
\section{Reaction rates for charmonium dissociation}\label{rates}

\begin{figure}[ht]
                 \insertplot{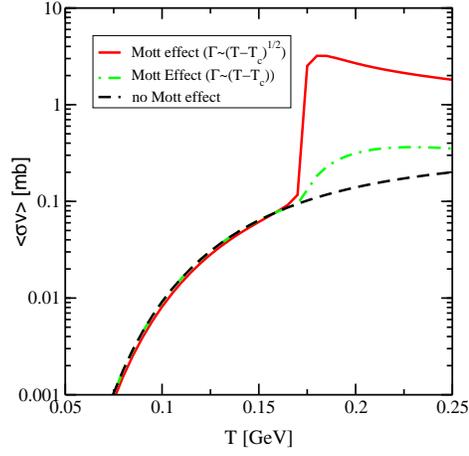}
\caption[]{Thermal average of the J/$\psi$ dissociation 
cross section without Mott
effect for the $D$- mesons (dashed line), with Mott effect and spectral 
function of Ref. \cite{bbk2} (dash-dotted line), and of Ref. \cite{bbvy} 
(solid line).}
\label{MottRateVGL_T}
\end{figure}

The inverse lifetime of a charmonium state due to collisions with 
light hadrons $h=\pi, \rho$ is given by 
$\tau^{-1}=\tau_\pi^{-1}+\tau_\rho^{-1}$ with \cite{bbk2}
\begin{eqnarray}
\tau^{-1}_h(T)&=&\int\frac{d^3p}{(2\pi)^3}\int ds' A_h(s';T)f_h(p,s';T)
j_h(p,s')\sigma_h^*(s;T)\nonumber\\
&=&\langle \sigma_h^* v\rangle n_h(T)~,
\end{eqnarray} 
where $f_h(p,s';T)$ is the Bose distribution function with the energy argument
$E(p,s')=\sqrt{p^2+s'}$,  $j_h(p,s)$ is the flux factor for $(c\bar c)$-$h$ 
collisions, and $\sigma_h^*(s;T)$ is the in-medium dissociation cross section  
\begin{equation}
\sigma_h^*(s;T)=\int~ds_1~ds_2~A_{D_1}(s_1;T)A_{D_2}(s_2;T)~\sigma_h(s;s_1,s_2)
~,
\end{equation}
which is displayed in Fig. \ref{sigmaT} for the spectral function fit of 
Ref. \cite{bbk2} as well as modified NJL model result of Fig. 
\ref{DmesonNJL}  (for $h=\pi$ we omit the subscript). 
In both cases the Mott effect for the $D$-meson final states 
entails a lowering of the threshold for the dissociation process. 
In Fig. \ref{MottRateVGL_T} we show that the $D$-meson Mott effect leads to a
strong enhancement in the thermal averaged dissociation 
cross section, i.e. in the
inverse lifetime of the J/$\psi$, which is quite sensitive to the details 
of the temperature dependence of the $D$-meson spectral function. 
This effect could be a key to understanding the physical mechanism
of anomalous J/$\psi$ suppression \cite{na50} and fast chemical 
equilibration \cite{pbm} in the CERN NA50 experiment. 
The role of the Mott effect in the 
$D$-$\bar D$ recombination rate ($D$-$\bar D$ fusion) 
merits further investigation. 

\section{Discussion of other processes}
\label{others}
In the QGP (and in the mixed phase), due to the presence of quasifree quarks 
and gluons, new channels for charmonium formation and dissociation exist which 
could drive chemical equilibration during the
existence of the fireball formed in the heavy-ion collision. 
This possibility has been considered within the statistical hadronization 
approach \cite{pbm,gorenstein} and seems to give appreciable contributions 
to J/$\psi$ production even under SPS conditions \cite{rapp}. 
A satisfactory description of the $\psi'/\psi$ ratio however requires an 
increase in the $\psi'$ dissociation 
rate, perhaps due to medium modification
of the $D$-meson threshold \cite{rapp} as provided for example 
by the Mott-effect
scenario in the present approach \cite{bbk2}. 
The role of partonic in-medium effects in charmonium kinetics in a QGP,
which has previously 
been discussed in the string-flip model of quark matter in the
form of a modified mass action law \cite{rbs} and dissociation rate 
\cite{rbs2}, should be reconsidered. 
We anticipate that rate coefficients for the ionization and recombination
of charm mesons could be described using an approach similar to methods
used previously to study
Coulomb plasmas \cite{rr,bornath}. 
   
\section{Conclusions}\label{concl}
A detailed description of quark substructure and $q\bar q$ wavefunctions is
is essential for the understanding of the behaviour of
meson-meson cross sections in the vacuum as well as modifications in
dense matter. We have shown that due to the Mott effect for $D$-mesons
at the QGP phase transition a reduction of the threshold for charmonium 
dissociation
occurs, which leads to a strong enhancement in the dissociation 
rate and a drop in 
the J/$\psi$ lifetime.
Important features of this approach are the off-mass-shell behaviour 
of the charmonium dissociation cross section and the 
hadronic spectral function, which we calculate using 
a modified NJL model.
The approximate agreement in scale we find between the
CLM and QIM dissociation cross sections must be considered
something of an accident,
since the QIM model cross section 
varies roughly according to the Povh-H\"ufner law for these processes,
whereas the CLM does not.
A unifying microscopic approach at the quark-gluon level can hopefully
be developed which will lead to additional insight into
the use of charm
(and bottom) production in heavy ion collisions as a diagnostic tool for
hot and dense matter.

 
\section*{Acknowledgements}
We are grateful to many colleagues for their discussions, in particular to 
P.-B. Gossiaux, J. H\"ufner, C.-M. Ko, S.H. Lee, Y. Oh, P. Petreczky, 
A. Polleri, R. Rapp and C.Y. Wong. 
The work of Yu.K. and G.B has been supported by DFG; D.B. and G.B. acknowledge 
support by DAAD for their visits at Oak Ridge National Laboratory and at the
University of Tennessee at Knoxville. T.B. and E.S.S. 
have been supported in part 
by NSF grant No. INT-0004089, and acknowledge the 
kind hospitality of the University of 
Rostock. D.B. thanks the organizers of the Workshop for creating
this wonderful meeting.
  

\begin{figure}[ht]
                 \insertplot{sigmaTlin.epsi} 
                 \insertplot{sigmaTnjl.epsi}
\caption[]{In-medium cross section for $J/\psi + \pi \to D + \bar D^*$ 
with the spectral function from \cite{bbk2} (upper plot) and from 
\cite{bbvy} (lower plot).}
\label{sigmaT}
\end{figure}
   
\end{document}